\documentstyle[preprint,12pt,aps,tighten]{revtex}

\def\mbpole{m_b^{\rm pole}}
\def\lambar{\bar\Lambda}
\def\mev{\,{\rm MeV}}
\def\gev{\,{\rm GeV}}
\def\d{{\rm d}}
\def\vub{|V_{ub}|}
\def\vcb{|V_{cb}|}

\begin{document}

{\preprint{\vbox{\hbox{JHU--TIPAC--97023}
      \hbox{hep-ph/9712364}
      \hbox{December 1997}}}

\title{Heavy Quark Effective Theory and Inclusive $B$ Decays}
\author{Adam F.~Falk}
\address{Department of Physics and Astronomy, The Johns Hopkins University\\
         3400 North Charles Street, Baltimore, Maryland 21218 USA\\
         {\tt falk@jhu.edu}}

\maketitle
\thispagestyle{empty}
\setcounter{page}{0}
\begin{abstract}

I review the use of the Heavy Quark Effective Theory in the computation of total
rates and differential kinematic distributions in inclusive semileptonic and
radiative $B$ decays.  Particular attention is paid to strategies for the
extraction of $V_{cb}$ and $V_{ub}$.

\end{abstract}

\vskip 3in

\centerline{\sl To appear in the Proceedings of Beauty '97}
\centerline{\sl University of California, Los Angeles}
\centerline{\sl October 12--16, 1997}

\newpage

\section{The heavy quark expansion for inclusive $B$ decays}

Why is it important to have a good theoretical understanding of inclusive weak
$B$ decays?  One might be tempted to take the point of view that since the
short distance structure of these decays is presumably known, all that is left
to do is to sort out some messy but ultimately unenlightening details of the QCD
dynamics.  However, by doing so one would forget the important fact that while
the structure of these decays is indeed well understood, the strengths of the weak
decay couplings of the $b$ quark are not yet known with sufficient precision.
The weak decay of the $b$ is governed by two fundamental constants which
appear as coefficients of currents found in the weak interaction Lagrangian.  The
larger of the two is $V_{cb}$, which is multiplies the current $\bar
c\gamma^\mu(1-\gamma^5)b$; the smaller is $V_{ub}$, which multiplies  $\bar
u\gamma^\mu(1-\gamma^5)b$.  Respectively, they govern the rates for the inclusive
semileptonic decays $B\to X_c\ell\nu$ and $B\to X_u\ell\nu$.  For reasons of
theoretical simplicity, I will confine myself to the discussion of
semileptonic, rather than nonleptonic, weak $B$ decays.\footnote{Actually, the
analysis of nonleptonic decays is on a less rigorous theoretical footing than
that of semileptonic decays~\cite{FWD}.}  In addition, I will discuss radiative
decays of the form $B\to X_{s,d}\,\gamma$.  These transitions, which are mediated
in the Standard Model by one-loop penguin operators, may also be studied with the
techniques to be described here.

The theoretical tool which is used to analyse semileptonic $B$ decays is the
Operator Product Expansion (OPE)~\cite{CGG,Bigi1}.  The OPE exploits a fact known
as parton-hadron duality, which is, roughly speaking, the observation that if the
energy release in a decay is large ($m_b\gg\Lambda_{\rm QCD}$) and the decay is
sufficiently inclusive, then it is essentially controlled by physics at short
distances.  Long distance effects, which arise from the $B$ meson bound state
structure, appear only as subleading corrections.  The result is an expansion for
the semileptonic width $\Gamma$ of the generic form
\begin{equation}\label{genwidth1}
  \Gamma=\Gamma_0\left[1+{\Lambda_1\over m_b}
  +{\Lambda^2_2\over m_b^2}+\dots\right],
\end{equation}
where $\Gamma_0$ is the decay rate of a free $b$ quark, and $\Lambda_i$ are
nonperturbative parameters of order $\Lambda_{\rm QCD}$ which are independent of
$m_b$ in the limit $m_b\to\infty$.  The quantities $\Lambda_i$ cannot be computed
theoretically from first principles (except, perhaps eventually, on the lattice);
instead, they must be modeled or, preferably, be measured.  It is the current
state of the art to include terms in $\Gamma$ up to order $1/m_b^3$.  Those of
order $1/m_b$ and $1/m_b^2$ are taken seriously, while those of order $1/m_b^3$
are used (usually crudely) to estimate the residual uncertainties, which are
typically at the level of a few percent.

Simultaneously, one may exploit the fact that the the QCD coupling constant
$\alpha_s$ is perturbative at the scale $m_b$, since $\alpha_s(m_b)\simeq0.2$. 
Since the typical gluon carries a momentum set by the scale $m_b\gg\Lambda_{\rm
QCD}$, the theory is in the region where asymptotic freedom applies.  As a
result, each term in the expansion (\ref{genwidth1}) is actually a power series
in $\alpha_s(m_b)$, leading to the refined (schematic) form
\begin{equation}\label{genwidth2}
  \Gamma=\Gamma_0\left[C_0(\alpha_s)+C_1(\alpha_s){\Lambda_1\over m_b}
  +C_2(\alpha_s){\Lambda^2_2\over m_b^2}+\dots\right].
\end{equation}
It is the present state of the art to compute $C_1$ to order $\alpha_s$ and
$C_0$ to ``order'' $\alpha_s^2\beta_0$, where $\beta_0=9$ is the first
coefficient in the QCD beta function.  Such terms, while not formally dominant in
any nearby limit, are often the largest contribution to the two loop
correction~\cite{BBZ}.  Here the largest uncertainties are from the choice of
renormalization scale $\mu$, and from infrared renormalons which appear, in
principle, at high orders in the perturbative expansion.  The renormalons will be
discussed further below.

To the order which they are computed, there are four input parameters which
appear in the expansion.  First, the QCD perturbation series depends on
$\alpha_s(m_b)$, which we will treat as known, since the uncertainty associated
with its value is comparatively small.  Second, there are parameters which arise
in the OPE.  The most important of these is $\lambar$, defined by
\begin{equation}
  \lambar=\lim_{m_b\to\infty}\left[M_B-\mbpole\right]\,.
\end{equation}
The constituent quark model, along with QCD sum rule estimates, lead one to
expect $200\mev\le\lambar\le700\mev$.  At the next order, there appear
two matrix elements of dimension five operators~\cite{FN},
\begin{eqnarray}
  \lambda_1&=&\langle B|\,\bar b(iD)^2b\,|B\rangle/2M_B\,,\nonumber\\
  \lambda_2&=&\langle B|\,
  \bar b\left(-{g\over2}\sigma^{\mu\nu}G_{\mu\nu}\right)b\,|B\rangle/6M_B\,.
\end{eqnarray}
The parameter $\lambda_1$ is related to the negative of the $b$ quark kinetic
energy in the $B$ meson; models and theoretical prejudice would indicate that it
lies in the range $0\le-\lambda_1\le1\gev^2$.  The parameter $\lambda_2$
is related to the matrix element of the leading operator which violates heavy
spin symmetry, and is proportional to the $B-B^*$ mass splitting.  Neglecting a
small radiative correction, $\lambda_2\approx0.12\gev^2$.  In what
follows, we will treat $\alpha_s$ and $\lambda_2$ as known, and $\lambar$
and $\lambda_1$ as parameters which must be determined somehow from experiment.

In fact, there is an additional subtlety associated with the definition of
$\lambar$, or more precisely, the definition of $\mbpole$.  The
problem is that $\mbpole$ is not well defined in QCD.  Of
course, nonperturbatively there is no pole in the fully dressed quark
propagator, hence no unambiguous definition of $\mbpole$.  But even within QCD
perturbation theory, the asymptotic nature of the expansion leads to an
ambiguity in the perturbative definition of $\mbpole$.  In particular, one can
attempt to sum the ``BLM-enhanced'' terms proportional to
$\alpha_s^n\beta_0^{n-1}$, where $\beta_0$ is the first coefficient in the QCD
beta function, using Borel resummation techniques.  However, one finds an
obstruction in the Borel plane due to an infrared renormalon ambiguity, which
must be resolved by the choice of a resummation scheme.  What this means,
essentially, is that $\mbpole$ has a scheme dependence in its definition, quite
analogous to the renormalization scale dependence of quantities defined in
dimensional regularization with $\overline{\rm MS}$ subtraction.  As with the
renormalization group, this scheme dependence cancels in physical quantities, if
the calculation is organized self-consistently.  The problem is not with the
predictivity of the theory, but with the physical interpretation of $\mbpole$. 
Since the renormalon ambiguity is of order $100\mev$~\cite{BBZ,NS,Bigi2}, there is
no ``preferred'' value for $\mbpole$ within a smaller precision than this.  The
most important practical point is that if the renormalon ambiguity is {\it not\/}
canceled consistently, then there is an {\it irreducible\/} uncertainty in the
input $\bar\Lambda$, which it inherits from $\mbpole$.  One of the issues which I
will address is the consistent treatment of this effect.

\section{Inclusive semileptonic $B$ decay}

I turn first to the extraction of $\vcb$ from the inclusive decay $B\to
X_c\ell\nu$.  The expression for the semileptonic width is written most
conveniently as a function of the ratio of spin averaged masses $\overline
M_D/\overline M_B$, where $\overline M_D=(M_D+3M_{D^*})/4$ and $\overline
M_B=(M_B+3M_{B^*})/4$.  These physical masses are, in turn, functions of the quark
masses $m_b$ and $m_c$ and the hadronic parameters $\lambar$, $\lambda_1$ and
$\lambda_2$, according to
\begin{eqnarray}
  M_B&=&\mbpole+\lambar-{\lambda_1+3\lambda_2\over2m_b}+\dots\,.\nonumber\\
  M_{B^*}&=&\mbpole+\lambar-{\lambda_1-\lambda_2\over2m_b}+\dots\,.
\end{eqnarray}
The result for the semileptonic width is~\cite{Bigi1,MW,Nir,LSW}
\begin{eqnarray}\label{vcbexp}
  \Gamma(B\to X_c\ell\nu)&=&{G_F^2\vcb^2 M_B^5\over192\pi^3}\,
  f\left(\overline M_D/\overline M_B\right)
  \left\{1-1.54{\alpha_s\over\pi}-(1.43\beta_0+c)\left(
  {\alpha_s\over\pi}\right)^2\right.\nonumber\\
  &&\left.\;\mbox{}-1.65{\lambar\over M_B}
  \left(1-0.87{\alpha_s\over\pi}\right)
  -0.95{\lambar^2\over M_B^2}-3.18{\lambda_1\over M_B^2}
  +0.02{\lambda_2\over M_B^2}+\dots\right\},
\end{eqnarray}
where $f(\overline M_D/\overline M_B)=0.369$ is a phase space factor, and the
two loop coefficient $c$ is known to be small~\cite{Czar}.  An important feature
of the QCD expansion is that there is also a renormalon ambiguity in the
perturbation series in Eq.~(\ref{vcbexp}); that is, there is a
scheme dependence in the summation of the series of terms proportional to
$\alpha_s^n\beta_0^{n-1}$.  However, one can show that this ambiguity precisely
compensates the renormalon ambiguity in $\lambar$, so the scheme dependence
cancels in the physical width~\cite{BBZ,Bigi2}.  The mechanism is similar to
the cancelation of renormalization scale dependence in physical quantities.

Once $\Gamma(B\to X_c\ell\nu)$ has been measured, the expression (\ref{vcbexp})
can be used to extract $\vcb$.  There is a variety of sources of uncertainty in
this determination.  First, both the perturbative QCD expansion and the operator
product expansion are truncated, and we must estimate the size of the omitted
terms.  For the radiative corrections, the leading unknown term is of order
$(\alpha_s/\pi)^2$, and is expected to be no larger than a few percent.  For the
nonperturbative corrections, the next terms are of order $(\lambar/M_B)^3$, also
at the level of a percent or so.  Second, there is uncertainty in the values of
the parameters to be inserted in the terms which have been calculated.  By far
the most important of these is the uncertainty in $\lambar$, or equivalently in
$\mbpole$, because it enters the expansion already at first order.  For example,
an uncertainty in $\lambar$ of $200\mev$ implies an uncertainty in $\vcb$ of
approximately 10\%.  This is the dominant source of error in the determination of
$\vcb$.

The cleanest and best way to reduce this uncertainty is to determine $\lambar$
directly from the semileptonic decays themselves.  This is possible because one
can compute not only total decay rates with the OPE, but kinematic distributions
as well.  While there exist interesting pathologies in certain of these
distributions, low moments of kinematic observables are almost always well
behaved.  In particular, it is useful to consider the following two types of
quantities:

$(i)$ In the decay $B\to X_c\ell\nu$, let $s_H$ be the invariant mass of the
hadronic state $X_c$.  Then the moments of the form $\langle (s_H-\overline
M_D^{\,2})^n\rangle$ have an OPE and a perturbative QCD expansion.  Constructed in
this way, the first moment starts at order $\lambar M_B$, while the second starts
at order $\lambar^2$.  Since terms of the order $\lambar^3$ are sources of
uncertainty, the first moment is substantially more reliable theoretically than
is the second.  These moments have been studied in detail in
Refs.~\cite{FLS,FL}}.

$(ii)$ One may also study the differential distribution $\d\Gamma/\d E_\ell$ and
its moments $\langle E_\ell^n\rangle$.  Again, these quantities have an OPE and a
perturbative QCD expansion.  They have been studied in detail in
Refs.~\cite{Gremm,Vol}.

Using a calculation analogous to that of the total semileptonic decay rate, these
kinematic moments have expansions in terms of $\alpha_s$, $\lambar$, $\lambda_1$
and $\lambda_2$.  As with the total rate, the renormalon ambiguity in the
perturbative series cancels that in $\lambar$.  Thus, if one writes $\Gamma(B\to
X_c\ell\nu)$ in terms of one of these moments, eliminating $\lambar$, the
renormalon ambiguity cancels automatically in the resulting expression.  For
example, if we eliminate $\lambar$ in favor of
$s_1\equiv\langle s_H-\overline M_D^2\rangle$ and solve for $\vcb$, we find
\begin{eqnarray}\label{vcbsolve}
  \vcb&=&
  \left[{\Gamma(B\to X_c\ell\nu)\over G_F^2M_B^5/192\pi^3}\right]^{-{1\over2}}
  \left\{1+0.59{\alpha_s\over\pi}+(0.37\beta_0+c')
  \left({\alpha_s\over\pi}\right)^2\right.\nonumber\\
  &&\left.\quad\mbox{}+0.68{s_1\over M_B^2}\left(1+0.70{\alpha_s\over\pi}\right)
  +0.38{s_1^2\over M_B^4}-2.00{\lambda_1\over M_B^2}+1.14{\lambda_2\over M_B^2}
  +\dots\right\}.
\end{eqnarray}
In this expression, the two loop contributions and the uncertainty due to
$\lambda_1$ are both at the level of one percent.  The effect of the leading
infrared renormalon has been canceled, and one expects the remaining
perturbation series in (\ref{vcbsolve}) to be much better behaved than in
the original expansion (\ref{vcbexp}).\footnote{An explicit example of how this
improvement works, up to order $\alpha_s^5$, is given in Ref.~\cite{FLS}.} Of
course, one then needs an accurate measurement of $s_1$ or another kinematic
moment.  In Ref.~\cite{FL}, it is shown that the theoretical sensitivity of $s_1$
and $\langle E_\ell\rangle$ to $\lambar$ are roughly the same, once corrections
of order $(\lambar/M_B)^3$ and necessary kinematic cuts are included.  Either
way, such an analysis is the strategy which will be required to reduce further the
experimental error of $\vcb$.

\section{Charmless inclusive semileptonic $B$ decay}

I turn now to charmless semileptonic $B$ decay, and the inclusive width
for $B\to X_u\ell\nu$.  The partial decay rate $\Gamma(B\to X_u\ell\nu)$ is
calculable as before, but unfortunately it is not very useful
phenomenologically.  The reason is that $\vub/\vcb\sim0.1$, so
approximately 99\% of semileptonic $B$ decays have charm in the final state. 
The only way to reject this huge background is by making kinematic cuts which
exclude charmed final states unambiguously.  

The simplest method experimentally, which has been used extensively in the past,
is to require that the charged lepton in the decay have an energy larger than
approximately $2.2\gev$, beyond the kinematic endpoint for decays to charm.  This
constitutes only a small tail of the spectrum $\d\Gamma/\d E_\ell$.  In order to
extrapolate to the complete partial width, it is necessary to have an accurate
theoretical understanding of the shape of $\d\Gamma/\d E_\ell$ in the endpoint
region.  Unfortunately, while the OPE can be used to predict $\d\Gamma/\d E_\ell$
over most of the available phase space, it is actually an expansion in powers of
$\lambar/(M_B-2E_\ell)$ rather than $\lambar/M_B$~\cite{MW,Neubert}.  In the
endpoint region, where $E_\ell\simeq M_B/2$, the OPE fails to converge reasonably
and cannot be used for an accurate analysis.

As a result, the extraction of $\vub$ by this method is extremely model
dependent.  Two approaches are used to address the issue.  The first is
inclusive, in which one performs the calculation as before but replaces the
higher order terms in the OPE with a ``shape function'', which is an ansatz which
resums formally an infinite set of terms in the series.  This shape function is
usually modeled by giving the $b$ quark a Gaussian momentum
distribution~\cite{ACCMM}.  The second is to treat the endpoint region as a sum
over decays to individual exclusive final states, such as $\pi\ell\nu$,
$\rho\ell\nu$, which are then analysed with a constituent quark potential
model~\cite{ISGW,BSW}.  

Each of these approaches has obvious deficiencies and misses important physics. 
Furthermore, even within each model there are uncertainties which cannot be
quantified or even estimated, since these models are in no sense systematic
approximations to QCD.  There is an important moral to be drawn.  In a situation
such as this, analysing additional models does not lead to better accuracy in the
determination of $\vub$.  Nor can the uncertainty in $\vub$ due to the use of
models be estimated sensibly simply by surveying the models currently available
on the market.  The model dependence in $\vub$ is impossible to quantify
reliably, and it continues to be underestimated consistently throughout the
literature on this subject.

However, with the development of techniques for neutrino reconstruction, a new
type of analysis has become possible.  If the energy and momentum of the missing
$\nu$ can be reconstructed kinematically, then it is possible to measure
inclusively the invariant mass $s_H$ of the final state $X_u$.  Charmed final
states then can be rejected by requiring $s_H<\Delta^2\le M_D^2$, where
$\Delta^2$ is an experimental cut which may have to be less than $M_D^2$ because
of details of the neutrino reconstruction procedure.  The advantage of this
approach is that most of the final states in $B\to X_u\ell\nu$ will pass such a
cut, if $\Delta^2$ is reasonably close to $M_D^2$.  One still must correct for the
small ``leakage'' of rate to larger values of $s_H$, but, in contrast to the
previous analysis, one is {\it not\/} extrapolating most of the rate from a small
tail.  Hence the larger $\Delta^2$ can be made to be, the smaller the effect of
this tail on the measurement of $\vub$, {\it even if the tail is not modeled
particularly well.}

A careful analysis~\cite{FLW} (see also Ref.~\cite{DU}) reveals that the
fraction of rate with $s_H>\Delta^2$ depends crucially on the value of
$\lambar$.  This is true for two reasons.  First, at tree level in the
parton model, the spectrum $\d\Gamma/\d s_H$ cuts off at $s_H=\lambar m_b$,
considerably below $M_D^2$.  Thus, gluon bremmstrahlung plays an important role
in creating states of large $s_H$, and this bremmstrahlung has a Sudakov-type
double logarithmic singularity at $s_H=\lambar m_b$.  The position of this
(integrable) singularity controls the strength of the tail of the spectrum. 
Second, the nonperturbative corrections from the OPE have the effect of smearing
the parton level spectrum by an amount of order $\lambar m_b$.  When this
smearing is modeled in the ACCMM model, it turns out that the dominant
sensitivity of the size of the tail is to the value of $\lambar$; once $\lambar$
is fixed, the dependence of the leakage on other parameters, such as $\lambda_1$,
is minimal.

Hence it is important, once again, to measure $\lambar$ using one of the
inclusive methods discussed in the previous section.  An accurate determination
of this parameter is crucial for the accurate determination of $\vub$.  In
addition, we should hope that the value of $\lambar$ turns out to be rather
small, and that the experimenters can make $\Delta^2$ close to $M_D^2$, so that
the separation between $\lambar m_b$ and $\Delta^2$ will be as large as
possible.  From the analysis of Ref.~\cite{FLW}, one may estimate the eventual
theoretical accuracy $\delta\vub$ under a variety of hypothetical scenarios: 
$(i)$ if $\Delta^2\simeq M_D^2$ and $\lambar=400\pm100\mev$, then
$\delta\vub\sim10\%$; $(ii)$ if $\Delta^2\simeq 1.5\gev^2$ and
$\lambar=400\pm100\mev$, then $\delta\vub\sim30\%$;
$(iii)$ if $\Delta^2\simeq 1.5\gev^2$ and $\lambar=200\pm100\mev$, then
$\delta\vub\sim10\%$; and $(iv)$ if $\Delta^2\simeq M_D^2$ and
$\lambar=600\pm100\mev$, then $\delta\vub\sim50\%$.  With luck, this method
eventually will yield the most accurate value of $\vub$ available, a value
largely free of model dependence.

\section{Radiative $B$ decay}

Finally, I turn to the extraction of information about short distance dynamics
from the inclusive radiative decay $B\to X_s\gamma$.  Away from the region in
which $M(X_s)\approx M(J/\psi)$, this decay is mediated primarily by a transition
magnetic moment operator of the form $\bar s\sigma^{\mu\nu}F_{\mu\nu}b$.  In
the Standard Model, this operator is induced by one loop GIM-violating
effects, and its coefficient is small.  The total branching fraction for
this decay in the Standard Model is at the level of $10^{-4}$.  One the other
hand, the coefficient of this operator, and hence the decay rate in this channel,
is substantially larger in many models of new physics, especially those that
postulate new flavor dynamics at the TeV scale.  In order for an observed
enhancement of $B\to X_s\gamma$ to serve as a signal of new physics, we must know
how to predict the physical inclusive rate from a known quark operator such as
$\bar s\sigma^{\mu\nu}F_{\mu\nu}b$.

In fact, the situation is quite analogous to that for $B\to X_u\ell\nu$, in that
the total rate for $B\to X_s\gamma$ can be computed reliably with the OPE but is
not quite useful phenomenologically~\cite{FLS2,Bigi3}.  Once again, the
difficulty is the much larger number of decays to charmed final states.  The most
straightforward way to search for $B\to X_s\gamma$ is to look for the relatively
hard $\gamma$, but there is an important background from the process $B\to
D\pi^0\to D\gamma\gamma$, where one of the photons is missed.  If one looks at
the photon energy spectrum $\d\Gamma/\d E_\gamma$, it is unambiguously from the
final state $X_s\gamma$ only for $E_\gamma>2.2\gev$.  As a result, one is again
in the position of extrapolating the bulk of the spectrum from a kinematic tail. 
For reasons similar to those which plague $B\to X_u\ell\nu$, the OPE is not
convergent in the region to which one is restricted.

However, in some respects the situation is not quite as severe as for $B\to
X_u\ell\nu$.  At tree level in the parton model, the photon is monochromatic
with $E_\gamma\simeq2.4\gev$, above the kinematic cut.  Radiative
corrections and nonperturbative bound state effects smear the spectrum about
this point, but most of the decay rate remains at large values of $E_\gamma$. 
In particular, even a relatively small downward shift in the value of $E_\gamma$
at which one must cut can reduce substantially the fraction of the rate
which must be inferred by extrapolation.  At present, however, the state of the
art is still to use an inclusive model in this analysis, based on an Gaussian
ansatz for the initial momentum distribution of the $b$ quark in the $B$
meson~\cite{AG}.

What is most important for the future is to aim for a measurement of $\d\Gamma/\d
E_\gamma$ down to values of $E_\gamma$ as small as possible.  Also, an accurate
measurement of the {\it shape\/} of the spectrum in the region $E_\gamma>2.2\gev$
would help substantially to reduce the model dependence of the result, by
providing information to constrain the models which are used.  An accurate
measurement of the spectrum in this region also could provide information on the
shape function which governs the endpoint of the lepton energy spectrum in $B\to
X_u\ell\nu$~\cite{Neubert,Bigi}, or could be used in an extraction of $\lambar$
and $\lambda_1$ which is complementary to that from the decay $B\to
X_c\ell\nu$~\cite{KL,Dikeman}.

\section{Summary}

Inclusive $B$ decays are indispensable for the accurate extraction of $\vcb$,
$\vub$, and the coefficient of the operator responsible for the decay $b\to
s\gamma$.  While the theoretical treatment of such decays has developed into a
mature field over the past five years, a number of subtleties remain.  For the
extraction of $\vcb$ from the dominant decay $B\to X_c\ell\nu$, the most important
theoretical uncertainties are those in the inputs $\lambar$ (or $m_b$) and (to a
lesser extent) $\lambda_1$, and from the truncation of the perturbative expansion
at order $\alpha_s^2$.  However, the infrared renormalon relates an ambiguity in
the definition of $\lambar$ to an ambiguity in the resummation of the
perturbative series, in such a way that the consistent elimination of $\lambar$
from physical quantities also helps to control the bad behavior of the
perturbation series.  Explicit strategies for extracting $\lambar$ from
experiment are discussed.

For the rare decays $B\to X_u\ell\nu$ and $B\to X_s\gamma$, more serious
theoretical uncertainties are induced by the severe kinematic cuts which must be
imposed to eliminate the background from $B\to X_c\ell\nu$.  A strategy for
extracting $\vub$ from $B\to X_u\ell\nu$, which relies on the experimental
technique of neutrino reconstruction, has been developed.  Future improvement in
the theoretical understanding of inclusive semileptonic and rare $B$ decays will
depend on $(i)$ the direct measurement of $\lambar$ and $\lambda_1$ in inclusive
decays themselves, a program which is already underway; and $(ii)$ application
of the loosest kinematic cuts possible in rare decays, which will help reduce the
model dependence of the results.  In the short and in the long term, the future in
this field is bright; ultimately, the prospects are excellent for obtaining
accurate and reliable insight into the CKM matrix and other short distance
physics from inclusive semileptonic and radiative $B$ decays.

\acknowledgments

I would like to thank the organizers for their gracious hospitality and a most
stimulating meeting.  This work was  supported in part by the National Science
Foundation under Grant No.~PHY-9404057 and under National Young Investigator
Award No.~PHY-9457916, by the Department of Energy under Outstanding Junior
Investigator Award No.~DE-FG02-94ER40869, and by the Alfred P.~Sloan Foundation.
A.F.~is a Cottrell Scholar of the Research Corporation.

\end{document}